# External luminescence and photon recycling in near-field thermophotovoltaics

John DeSutter[*], Rodolphe Vaillon[**,*] and Mathieu Francoeur[*,†]

[*]*Radiative Energy Transfer Lab, Department of Mechanical Engineering, University of Utah, Salt Lake City, UT 84112, USA*

[**]*Univ Lyon, CNRS, INSA-Lyon, Université Claude Bernard Lyon 1, CETHIL UMR5008, F-69621, Villeurbanne, France*

**ABSTRACT**

The importance of considering near-field effects on photon recycling and spontaneous emission in a thermophotovoltaic device is investigated. Fluctuational electrodynamics is used to calculate external luminescence from a photovoltaic cell as a function of emitter type, vacuum gap thickness between emitter and cell, and cell thickness. The observed changes in external luminescence suggest strong modifications of photon recycling caused by the presence of the emitter. Photon recycling for propagating modes is affected by reflection at the vacuum-emitter interface and is substantially decreased by the leakage towards the emitter through tunneling of frustrated modes. In addition, spontaneous emission by the cell can be strongly enhanced by the presence of an emitter supporting surface polariton modes. It follows that using a radiative recombination model with a spatially uniform radiative lifetime, even corrected by a photon recycling factor, is inappropriate. Applying the principles of detailed balance, and accounting for non-radiative recombination mechanisms, the impact of external luminescence enhancement in the near field on thermophotovoltaic performance is investigated. It is shown that unlike isolated cells, the external luminescence efficiency is not solely dependent on cell quality, but significantly increases as the vacuum gap thickness decreases below 400 nm for the case of an intrinsic silicon emitter. In turn, the open-circuit voltage and power density benefit from this enhanced external luminescence toward the emitter. This benefit is larger as cell quality, characterized by the contribution of non-radiative recombination, decreases.

---

[†] Corresponding author. Tel.: + 1 801 581 5721
Email address: mfrancoeur@mech.utah.edu

# I. INTRODUCTION

A thermophotovoltaic (TPV) device converts thermal energy into electricity and consists primarily of an emitter and a photovoltaic (PV) cell separated by a vacuum gap [1,2]. The emitter is heated by an external source such as the sun [3,4] or a residential boiler [5] among others and, in turn, emits thermal radiation which is used to generate electron-hole pairs (EHPs) within the PV cell. TPV devices hold great potential for both solar-to-electrical energy conversion and waste heat recovery, and are expected to be quiet, modular, safe and pollution free [1].

TPV performance can be enhanced when the vacuum gap separating the emitter and cell is decreased to less than the thermal wavelength $\lambda_T$. At a sub-wavelength gap thickness, thermal radiation is in the near-field regime such that heat transfer can exceed the blackbody limit. This is due to the contribution of evanescent modes, decaying exponentially within a distance of approximately a wavelength normal to the surface of a heat source, which accompany the propagating modes described by Planck's theory [6-17]. These evanescent modes can be generated by total internal reflection at an interface (frustrated modes) and through phonon or free electron oscillations (surface polariton modes). In polar dielectric materials such as silicon carbide, surface phonon-polaritons are generated by the out of phase oscillations of transverse optical phonons [7]. The out of phase longitudinal free electron oscillations in metals and doped semiconductors lead to surface plasmon-polaritons [8]. In a TPV system, it is expected that the contribution of evanescent modes leading to an increase of photon absorption, and thus an increase of the generation rate of EHPs in the cell, results in enhanced device performance [18-35].

In addition to the enhancement of photon absorption in the near field, there is also an increase in photon emission from the cell lost to the surroundings, or external luminescence. This external luminescence enhancement can be appropriately accounted for [24-26,31,33] using the principles of detailed balance [36]. Yet, the physics behind this increase as it relates to cell radiative recombination of EHPs is not well understood. Indeed, an approach commonly used to model radiative recombination in near-field TPVs involves a spatially uniform radiative lifetime in which only spontaneous photon emission, or internal luminescence, in the cell is considered [21,23,28,30,32,34]. It has been shown that with this approach, the reabsorption of internally emitted photons, or photon recycling, is neglected [37]. Radiative lifetime can be corrected using a spatially uniform photon recycling factor, but this does not account for near-field impacts on photon recycling. Modeling radiative recombination via a spatially uniform radiative lifetime also neglects potential near-field impacts on the internal spontaneous emission within cell [21,38]. Laroche et al. [21] concluded that the impact of surface polariton modes on radiative lifetime is negligible, since it only affects a very small portion of the cell. However, photon recycling was not discussed in their analysis.

In this work, the near-field impacts on photon recycling and spontaneous emission as they relate to radiative recombination in TPV devices are investigated. This is accomplished by analyzing the different photon modes contributing to the cell near-field external luminescence calculated via fluctuational electrodynamics. The influence on external luminescence of the presence of the emitter, emitter type, vacuum gap thickness and cell thickness are investigated. Internal and external luminescence of an isolated cell, i.e. with no emitter, are used to evaluate the modification in photon recycling caused by the presence of the emitter. In addition, the impact of enhanced external luminescence on near-field TPV performance is analyzed when including non-radiative recombination mechanisms.

## II. DESCRIPTION OF THE PROBLEM

The near-field TPV device outlined in Fig. 1 is considered, in which a semi-infinite bulk emitter and a PV cell with thickness $t$ are separated by a vacuum gap of thickness $d$. The emitter and the cell are at constant and uniform temperatures of $T_e$ = 800 K and $T_c$ = 300 K. A temperature of 800 K is chosen for the emitter as it is a representative value for waste heat [31]. The cell consists of gallium antimonide (GaSb) and has a bandgap energy of $E_g$ = 0.72 eV (bandgap frequency of $\omega_g$ = 1.09×10$^{15}$ rad/s) at 300 K [39]. GaSb is chosen since it is a well-established cell material that can be fabricated at lower cost than materials with smaller bandgaps as epitaxial processes are not required [40]. For frequencies above $\omega_g$, the interband dielectric function of GaSb is calculated via the model provided in Ref. [41] and is assumed to be independent of dopant level and type. In this frequency band, the lattice and free carriers contribution to the dielectric function is negligible [23]. The cell substrate is modeled as a semi-infinite layer with a frequency independent dielectric function of $\varepsilon_s$ = 1, thus corresponding to vacuum. This choice is made for purpose of comparison against the case of an isolated cell surrounded by vacuum and does not compromise the main conclusions of the analysis. Two types of emitters are considered: an emitter made of intrinsic silicon (Si) where the dielectric function is taken from Ref. [42] and an emitter made of a material supporting surface polariton modes. The dielectric function of the material supporting surface polariton modes is described by a Drude model $\varepsilon_{Dr} = 1 - \omega_p^2/(\omega^2 + i\omega\Gamma)$, where $\omega_p$ is the plasma frequency and $\Gamma$ is the loss coefficient [43]. Radiatively-optimized values of $\omega_p$ and $\Gamma$, fixed at 1.83×10$^{15}$ rad/s and $2.10\times10^{13}$ rad/s, respectively, are chosen to maximize above bandgap radiation transfer between the emitter and the cell [28]. These values lead to surface polariton resonance at 0.85 eV when sharing an interface with vacuum. This is similar to real materials such as titanium carbide (TiC) and tantalum silicide (TaSi$_2$) which have resonances at 0.9 eV and 0.8 eV, respectively [33]. The device is assumed to be azimuthally symmetric and infinite in the $\rho$-direction making the view factor unity between the emitter and the cell. This implies that only variations of absorption and emission in the $z$-direction are of concern. The cell is discretized strictly along $z$ into $N$ discrete layers of equal thickness $\Delta z_j$.

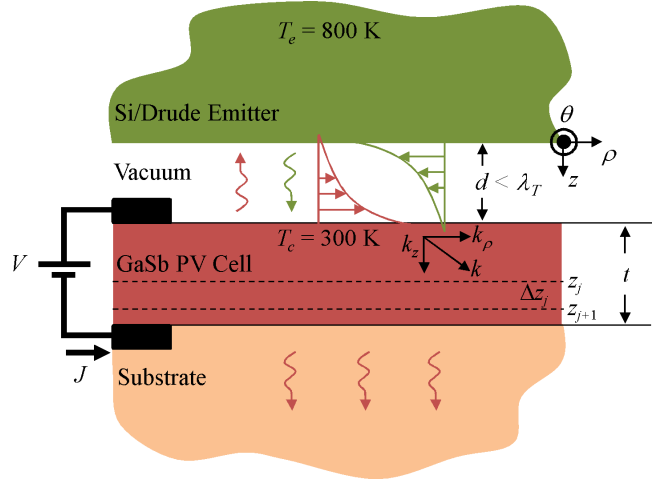

FIG. 1. (Color online) Schematic of a near-field TPV device showing contributions of propagating and evanescent modes to photon exchange. The emitter and PV cell, separated by a vacuum gap of thickness ($d$) smaller than the thermal wavelength ($\lambda_T$), are at constant and uniform temperatures of 800 K and 300 K, respectively.

The available photon modes contributing to external luminescence from the cell can be described in terms of the parallel wavevector $k_\rho$ [rad/m] (Fig. 1). Propagating modes in the cell with frequency $\omega$ are characterized by parallel wavevectors $0 < k_\rho < \mathrm{Re}(m_c)k_0$, where $m_c$ is the refractive index of the cell and $k_0 = \omega/c_0$ is the vacuum wavevector. To better understand how the presence of the emitter impacts the available photon modes contributing to external luminescence, an isolated cell surrounded by vacuum is first considered. In this case, only propagating modes in vacuum with $0 < k_\rho < k_0$ can escape the cell and contribute to external luminescence. These photon modes can be partially recycled, a phenomenon through which a photon generated via EHP recombination is reabsorbed by the cell and generates an additional EHP [37]. Here, photon recycling includes partial reabsorption of internally emitted modes along their pathlength. Photon modes with $k_0 < k_\rho < \text{Re}(m_c)k_0$ cannot escape the cell due to total internal reflection at the boundaries and, therefore, are all recycled.

In a TPV device where the emitter is in the far field of the cell ($d \gg \lambda_T$), external luminescence is gap-independent and is limited to propagating modes in vacuum. In addition to the mechanisms outlined above for the isolated cell, photon recycling can occur via reabsorption by the cell of modes reflected at the vacuum-emitter interface.

In a TPV device where the emitter and the cell are separated by a sub-wavelength gap ($d < \lambda_T$), external luminescence is gap-dependent due to interference of propagating modes and tunneling of evanescent modes from the cell to the emitter. Propagating modes in the cell experiencing total internal reflection generate frustrated modes with decaying evanescent fields in vacuum. Tunneling of these frustrated modes from the cell to the emitter increases the available photon modes that can escape the cell beyond the vacuum limit of $k_0$, thus decreasing the number of recycled photons compared to the far-field case. If, instead, a Drude emitter supporting surface polariton modes is used, the available $k_\rho$ modes can be further increased beyond the propagating limit in the cell, Re($m_c$)$k_0$, as their presence can greatly modify the local density of photon modes in the cell [38], impacting spontaneous emission.

The near-field external luminescence of the GaSb cell in the presence of the Si and Drude emitters is analyzed next.

## III. NEAR-FIELD EXTERNAL LUMINESCENCE

The current density $J$ [A/cm$^2$] in the cell can be described by the difference between the rate of above bandgap photon absorption per unit area $\gamma_a$ [(photons)/(cm$^2$s)] and the rate of above bandgap photon emission lost to the surroundings per unit area, also called external luminescence, $\gamma_e$ [(photons)/(cm$^2$s)] [36]:

$$J(d,V) = q[\gamma_a(d) - \gamma_e(d,V)] \tag{1}$$

where $V$ [V] is the applied voltage of an external load and $q$ [1.6022×10$^{-19}$ C] is the electron charge. The distinction of emission lost to the surroundings is made as in the presence of an emitter, some of the cell emission can be reflected at the vacuum-emitter interface and reabsorbed by the cell, thus contributing to photon recycling. Equation (1) is valid assuming that there is only radiative recombination (radiative limit), every above bandgap photon absorbed by the cell generates one EHP, every EHP that recombines produces one photon and charge carriers have infinite mobility allowing for all generated electrons and holes to be collected [36,37,44]. The main benefit of near-field TPVs is the enhancement of photon absorption $\gamma_a$ due to tunneling of evanescent modes from the emitter to the cell.

To appropriately account for all photon modes described in section II, the fluctuational electrodynamics formalism is used to calculate radiative exchange between the emitter and cell.

It involves the addition to Maxwell's equations of a thermally induced fluctuating current density representing thermal emission [45]. The link between the local temperature of a heat source and the fluctuating current density is provided by the fluctuation-dissipation theorem [45]. Fluctuational electrodynamics is applicable for gap thicknesses in both the far and near field, and its validity has been confirmed experimentally [46] and theoretically [47] down to separation gaps of 2 nm and 1 nm, respectively. Using this formalism, the above bandgap photon flux absorbed by the cell is calculated by summing the rate of photons absorbed within a discrete layer $\Delta z_j$ over all $N$ layers [31]:

$$\gamma_a(d) = \sum_{j=1}^{N} \int_{\omega_g}^{\infty} \frac{1}{\hbar\omega} \Theta(\omega, T_e, 0) \, \Phi_{e-c}(\omega, d, \Delta z_j) \, d\omega \qquad (2)$$

where $\Theta$ is the mean energy of a generalized Planck oscillator defined as [31,48]:

$$\Theta(\omega, T, V) = \frac{\hbar\omega}{\exp\left[(\hbar\omega - qV)/k_b T\right] - 1} \qquad (3)$$

The term $\Phi_{e-c}$ is the spectral, gap-dependent transmission factor relating the emitter to a discrete layer $\Delta z_j$ within the cell. The transmission factor, provided in the Appendix, is calculated from dyadic Green's functions and accounts for all modes, propagating and evanescent [49].

External luminescence, which accounts for emission lost toward the emitter ($\gamma_{e,c-e}$) and the substrate ($\gamma_{e,c-s}$), is given by:

$$\gamma_e(d, V) = \sum_{j=1}^{N} \int_{\omega_g}^{\infty} \frac{1}{\hbar\omega} \Theta(\omega, T_c, V) \left[\Phi_{c-e}(\omega, d, \Delta z_j) + \Phi_{c-s}(\omega, d, \Delta z_j)\right] d\omega \qquad (4)$$

where $\Phi_{c-e}$ and $\Phi_{c-s}$ are the spectral, gap-dependent transmission factors relating a discrete layer $\Delta z_j$ within the cell to the emitter and substrate, respectively. The transmission factor $\Phi_{c-e}$ is the same as $\Phi_{e-c}$ due to the reciprocity of the dyadic Green's functions [23]. The transmission factor $\Phi_{c-s}$ is provided in the Appendix. Equation (4) is valid for non-degenerate conditions and when the charge carriers have infinite mobility thus allowing for uniform quasi-Fermi level splitting throughout the cell which can be described by the applied voltage as $qV$ [37,44].

Under the Boltzmann approximation, which is appropriate when $(E_g - qV) \gg k_bT_c$ [44,50] and when $E_g$ is larger than 0.5 eV [50], external luminescence can be expressed as [37]:

$$\gamma_e(d,V) = \gamma_e^0(d)\exp\left[\frac{qV}{k_bT_c}\right] \quad (5)$$

In Eq. (5), the contribution in chemical equilibrium ($V = 0$) to external luminescence $\gamma_e^0$ [(photons)/(cm$^2$s)] is:

$$\gamma_e^0(d) = \sum_{j=1}^N \int_{\omega_g}^\infty \frac{1}{\exp[\hbar\omega/k_bT_c]}\left[\Phi_{c-e}(\omega,d,\Delta z_j) + \Phi_{c-s}(\omega,d,\Delta z_j)\right]d\omega \quad (6)$$

Figure 2 shows the cell external luminescence in chemical equilibrium per unit angular frequency and parallel wavevector, $\gamma_{e,\omega,k_\rho}^0$, for the Si and Drude emitters when the gap and cell thicknesses are fixed at $d = 10$ nm and $t = 10$ μm. In the presence of the Si emitter, frustrated modes described by normalized parallel wavevectors $1 < k_\rho/k_0 < \min[\text{Re}(m_{Si}), \text{Re}(m_c)]$, where $m_{Si}$ is the refractive index of Si, contribute significantly to the cell near-field external luminescence. For the case of the GaSb cell, the upper $k_\rho/k_0$ limit is $\text{Re}(m_{Si})$ for all frequencies of interest. Note that Re($m_{Si}$) = 3.46 at the bandgap frequency $\omega_g$ of GaSb, and is nearly constant in the $\omega$-band shown in Fig. 2(a). This implies that the presence of a Si emitter in the near field of the cell opens a new channel for external luminescence since modes with $1 < k_\rho/k_0 < $ Re($m_{Si}$), fully contributing to photon recycling in far-field TPVs, can now escape the cell. This has the effect of decreasing photon recycling and, consequently, increasing the impact of radiative recombination. Modes with Re($m_{Si}$) < $k_\rho/k_0$ < Re($m_c$) cannot propagate in Si and, therefore, cannot be tunneled from the cell to the emitter and are thus completely recycled. External luminescence towards the substrate, occurring through modes described by $0 < k_\rho/k_0 < 1$, remains unchanged from the case of an isolated cell causing the sudden transition at $k_\rho/k_0 = 1$.

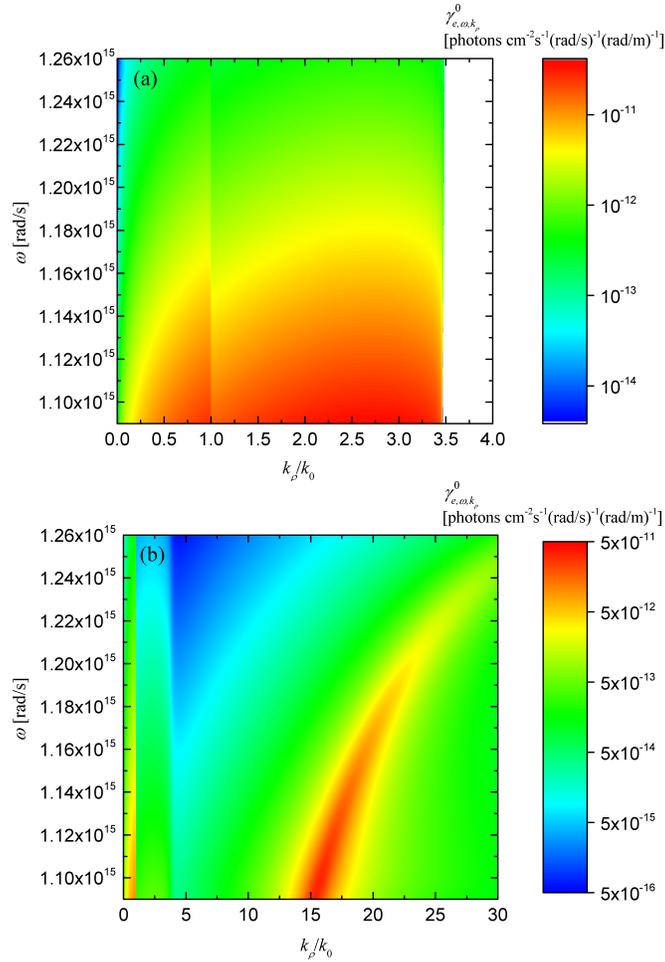
11

FIG. 2. (Color online) Cell external luminescence in chemical equilibrium ($\gamma^0_{e,\omega,k_\rho}$) as a function of angular frequency ($\omega$) and normalized parallel wavevector ($k_\rho/k_0$) for $d = 10$ nm and $t = 10$ µm: (a) Si emitter. (b) Drude emitter.

The photon mode limit established for the Si emitter is not appropriate when the emitter supports surface polariton modes. This is the case for the Drude emitter in which there is a significant enhancement in the available photon modes beyond the cell propagating limit of $k_\rho/k_0 = \text{Re}(m_c)$ contributing to external luminescence (Fig. 2(b)). This is because the local density of photon modes in the cell is greatly modified by the presence of the Drude emitter when $d < \lambda_T$ [38]. At the frequencies where the Drude emitter supports surface polaritons, the limiting mode contributing to external luminescence is gap-dependent and can be approximated by $k_\rho \approx 1/d$ [51,52] which leads to a value of $k_\rho/k_0 \approx 27$. This is in reasonable agreement with the results in Fig. 2(b). Therefore, an emitter supporting surface polariton modes opens an additional channel for external luminescence. Here, since the Drude emitter supports surface polaritons at frequencies for which thermal emission by the cell is significant, this leads to a large enhancement of the cell external luminescence.

In Fig. 3, external luminescence in chemical equilibrium, $\gamma^0_e$, is shown as a function of vacuum gap thickness $d$ for the Si and Drude emitters when the cell thickness is fixed at 10 µm. In order to quantify the modification of photon recycling due to the presence of the emitter, internal luminescence is used as a reference. Volumetric internal luminescence within an isolated PV cell in chemical equilibrium is quantified by the rate of above bandgap spontaneous photon emission per unit volume $\bar{\gamma}^0_{i,ic}$ [(photons)/(cm$^3$s)] [53,54]:

$$\bar{\gamma}_{i,ic}^{0} = \frac{1}{\pi^{2}c_{0}^{2}} \int_{\omega_{g}}^{\infty} (\text{Re}(m_{c}))^{2} \alpha_{c} \frac{\omega^{2}}{\exp[\hbar\omega/k_{b}T_{c}]-1} d\omega \tag{7}$$

where $\alpha_c$ [cm$^{-1}$] is the frequency-dependent absorption coefficient of the cell calculated using the dielectric function of GaSb. Under the Boltzmann approximation, the minus one term in the denominator of Eq. (7) can be removed such that under a bias $V$, the rate of spontaneous photon emission cumulated over the cell is given by [48,54]:

$$\gamma_{i,ic}(V) = \gamma_{i,ic}^{0} \exp\left[\frac{qV}{k_{b}T_{c}}\right] \tag{8}$$

where $\gamma_{i,ic}^{0} = \bar{\gamma}_{i,ic}^{0} t$, assuming that the temperature and radiative properties are uniform throughout the cell. Internal luminescence $\gamma_{i,ic}^{0}$ and external luminescence $\gamma_{e,ic}^{0}$ of an isolated cell are plotted in Fig. 3 as references. The difference between these two quantities indicates the photon recycling level for an isolated cell due to reflection at its boundaries and reabsorption within the cell.

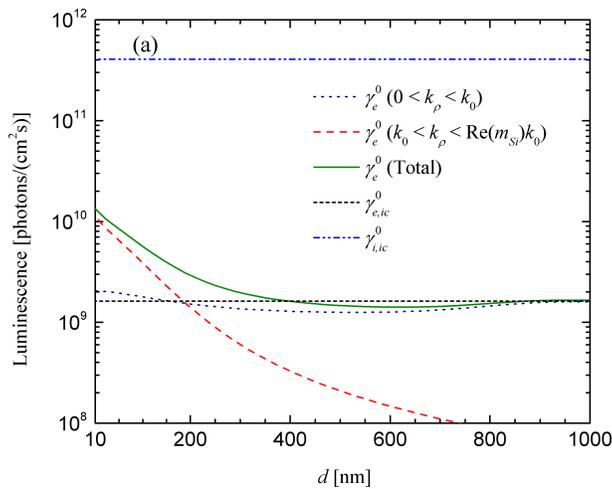

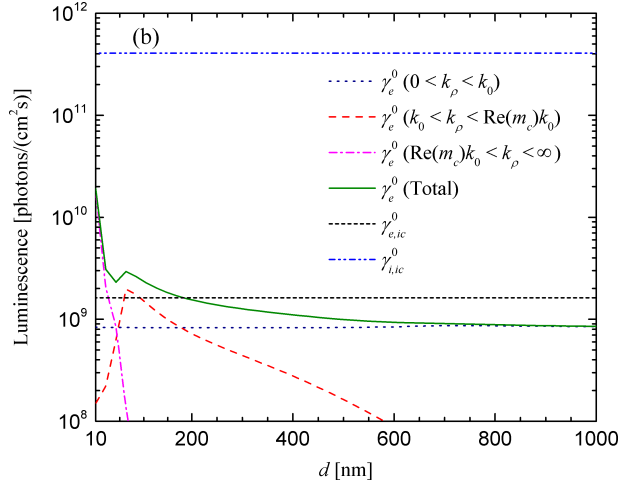

FIG. 3. (Color online) Cell external luminescence in chemical equilibrium ($\gamma_e^0$) as a function of the gap thickness ($d$) for $t = 10$ μm: (a) Si emitter. (b) Drude emitter. For comparison, internal luminescence ($\gamma_{i,ic}^0$) and external luminescence ($\gamma_{e,ic}^0$) of an isolated cell are also plotted.

For the case of the Si emitter (Fig. 3(a)), external luminescence of propagating modes in vacuum ($0 < k_\rho < k_0$) varies with gap thickness. This is due to coherence effects arising from multiple reflections in the vacuum gap [55,56]. External luminescence is up to 25% smaller or larger than the case of an isolated cell. For the Drude emitter (Fig. 3(b)), the impact from propagating modes in vacuum is reduced by a factor of approximately two compared to the case of an isolated cell since a significant portion of emission towards the emitter is reflected back to the cell and recycled. Thus, the corresponding external luminescence is dominated by emission towards the substrate.

The contribution to external luminescence by evanescent modes (frustrated and surface polariton modes) with $k_\rho > k_0$ is a strong function of the gap thickness, and becomes dominant at gap thicknesses smaller than 200 nm for both emitters. In the presence of the Si emitter, external luminescence increases (photon recycling decreases) by factors of 1.1 and 9.1 with respect to the far-field value at gap thicknesses of 1000 nm and 10 nm, respectively. In the limit that $d \rightarrow 0$, the cell external luminescence saturates since it is limited by modes with parallel wavevector $k_\rho < \text{Re}(m_{Si})k_0$. The Drude emitter allows tunneling all frustrated modes (propagating in the cell with $k_0 < k_\rho < \text{Re}(m_c)k_0$) dominating external luminescence between gap thicknesses of approximately 200 nm down to 50 nm. For gap thicknesses smaller than 50 nm, surface polariton modes in the cell with $k_\rho > \text{Re}(m_c)k_0$ dominate external luminescence. This change of dominant mode is caused by a shift in the dispersion relation at the emitter-vacuum interface due to the presence of the cell when the gap thickness is smaller than 70 nm. This shift was explained in Refs. [21,28] and reduces the contribution from modes $k_0 < k_\rho < \text{Re}(m_c)k_0$ and increases that from modes with parallel wavevectors exceeding $\text{Re}(m_c)k_0$. In the limit that $d \rightarrow 0$, external luminescence does not saturate but rather diverges since the limiting mode is characterized by a wavevector inversely proportional to the gap size ($k_\rho \approx 1/d$). At gap thicknesses of 10 nm and 1000 nm, the cell external luminescence respectively increases by a factor of 22.6 and remains unchanged with respect to the far-field value. Clearly, while near-field TPV devices have the benefit of increasing radiation absorption by the cell due to tunneling of evanescent modes, the cell external luminescence is also substantially enhanced. The external luminescence enhancement is due to a drop in photon recycling, caused by the leakage towards the emitter of frustrated modes, and the intensification of spontaneous emission when the emitter supports surface polariton modes.

The volumetric contribution by each discrete layer to external luminescence [(photons)/(cm$^3$ s)] is plotted in Fig. 4 as a function of cell depth for the Si and Drude emitters and for gap and cell thicknesses of $d = 10$ nm and $t = 10$ µm. This quantity is derived from Eq. (6) as:

$$\bar{\gamma}_e^0(d,\Delta z_j) = \frac{1}{\Delta z_j} \int_{\omega_g}^{\infty} \frac{1}{\exp[\hbar\omega/k_b T_c]} \left[ \Phi_{c-e}(\omega,d,\Delta z_j) + \Phi_{c-s}(\omega,d,\Delta z_j) \right] \tag{9}$$

The volumetric internal luminescence of an isolated cell, $\bar{\gamma}_{i,ic}^0$, and the local volumetric contribution to external luminescence of an isolated cell, $\bar{\gamma}_{e,ic}^0$, are also plotted in Fig. 4. The former is uniform within the cell and the latter is symmetric with respect to the center of the cell. In the presence of an emitter, this symmetry is broken because of the addition of evanescent modes to external luminescence towards the emitter. The spatial distribution is a strong function of the emitter type. Local contributions to external luminescence are largest everywhere in the presence of the Si emitter except near the front surface of the cell (i.e., $z = 0$). Surface polaritons supported by the Drude emitter greatly enhance the cell local density of photon modes near $z = 0$. This is because the penetration depth in the cell of surface polariton modes dominating radiative transfer is approximately equal to the gap size $d$ [52]. Near the cell front surface, the contribution to external luminescence with the Drude emitter even exceeds the internal luminescence of an isolated cell by over an order of magnitude. The internal luminescence of an isolated cell, which does not account for the impact on spontaneous emission of surface polariton modes supported by the Drude emitter [21], is clearly invalid.

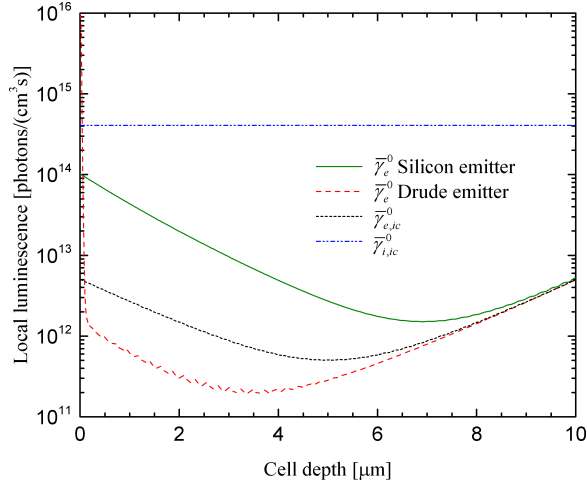

FIG. 4. (Color online) Local volumetric contribution to the cell external luminescence in chemical equilibrium ($\bar{\gamma}_e^0$) as a function of depth into the cell for $d$ = 10 nm and $t$ = 10 μm. For comparison, local volumetric internal luminescence in chemical equilibrium ($\bar{\gamma}_{i,ic}^0$) and local volumetric contribution to external luminescence of an isolated cell ($\bar{\gamma}_{e,ic}^0$) are plotted.

The external luminescence $\gamma_e^0$ dependence on cell thickness is shown in Fig. 5 for the Si and Drude emitters and for a gap thickness of $d$ = 10 nm. The curves are compared against internal luminescence of an isolated cell $\gamma_{i,ic}^0$. Regardless of cell thickness, external luminescence with the Si emitter is smaller than internal luminescence because of photon recycling. The same is observed with the Drude emitter down to a cell thickness of 0.46 μm. As the cell thickness decreases, an increasing portion of its volume is affected by the presence of the Drude emitter. This implies that a larger portion of the cell can spontaneously emit modes with parallel wavevector exceeding Re($m_c$)$k_0$ that are not taken into account in the internal luminescence model. In the limit that the cell thickness is comparable to the penetration depth of surface polaritons, which is approximately equal to the gap thickness, the density of modes of the entire cell is modified by the presence of the Drude emitter. For a 100-nm-thick cell, the external luminescence in the presence of the Drude emitter exceeds the internal luminescence of an isolated cell by nearly a factor of three. This means that spontaneous emission, and therefore radiative lifetime, can be greatly modified by the presence of an emitter supporting surface polariton modes, especially if the cell thickness is comparable to the vacuum gap thickness.

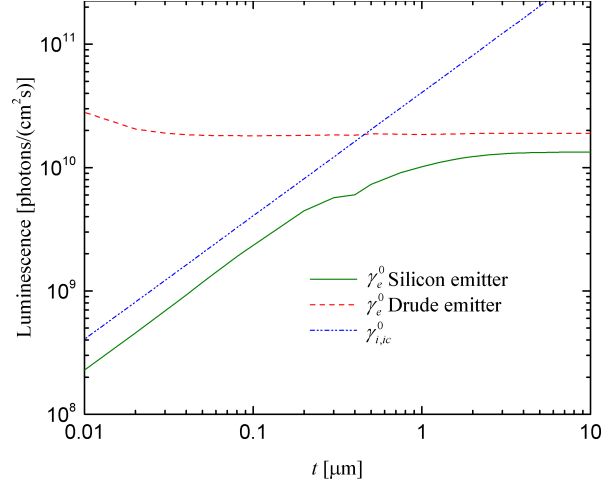

FIG. 5. (Color online) Cell external luminescence in chemical equilibrium ($\gamma_e^0$) as a function of the cell thickness ($t$) for $d = 10$ nm. For comparison, internal luminescence of an isolated cell ($\gamma_{i,ic}^0$) is plotted.

From Figs. 3 to 5 it is clear that internal luminescence of an isolated cell inaccurately represents the impact of radiative recombination in near-field TPVs. In determining the internal luminescence of an isolated cell, photon recycling and any increase in available modes beyond Re($m_c$)$k_0$ due to the presence of the Drude emitter are neglected. Despite this, internal luminescence of an isolated cell is often used to account for radiative recombination in near-field TPV devices through the implementation of a spatially uniform radiative lifetime $\tau_{rad}$ [s] [21,23,28,30,32,34]. In low injection conditions, radiative lifetime of an isolated cell can be defined by $\tau_{rad} = 1/BN$, where $N$ [carriers/cm$^3$] is the doping concentration and $B$ [(photons)cm$^3$/s] is the radiative recombination coefficient. This coefficient is related to internal luminescence of an isolated cell via the relation $B = \bar{\gamma}_{i,ic}^0 / n_i^2$, where $n_i$ [carriers/cm$^3$] is the intrinsic carrier concentration [38,57]. Near-field effects on photon recycling and spontaneous emission are, therefore, neglected when using a spatially uniform radiative lifetime $\tau_{rad}$.

In certain previous near-field TPV analyses, a spatially uniform photon recycling factor $\phi$ of 10 was used for GaSb to correct the radiative lifetime $\tau_{rad}$ to account for photon recycling [23,28,30,34]. Such a factor is somehow arbitrary and neglects the aforementioned near-field impacts on photon recycling and spontaneous emission. A consistent photon recycling factor would need to be dependent on the emitter type, cell and gap thicknesses as well as location in the cell (see Figs. 3 to 5). For example, for a 10-nm-thick gap and a cell thickness of 10 µm, the photon recycling factor $\phi$ at the cell surface is 4.03 and 0.04 for Si and Drude emitters, respectively. A $\phi$ value less than one implies that external luminescence exceeds internal luminescence of an isolated cell. These photon recycling factors for the Si and Drude emitters increase to 149.40 and 1433.35, respectively, at a cell depth of 5 µm. It is clear that modeling radiative recombination with a spatially uniform radiative lifetime $\tau_{rad}$ corrected by a spatially uniform photon recycling factor $\phi$ should be avoided in near-field TPV performance prediction. Instead, a full radiation model, i.e. fluctuational electrodynamics with the generalized Planck function, should be used.

The impact of external luminescence enhancement in the near field on TPV performance is analyzed next when considering non-radiative recombination mechanisms.

# IV. IMPACT OF EXTERNAL LUMINESCENCE ON NEAR-FIELD TPV PERFORMANCE

Assuming low injection conditions, non-radiative bulk Auger and Shockley-Read-Hall (SRH) recombination mechanisms are added to Eq. (1) as follows [31,38]:

$$J(d,V) = q\left[\gamma_a(d) - \gamma_e(d,V) - U_{Aug}(V) - U_{SRH}(V)\right]$$
$$= q\left[\gamma_a(d) - \left(\gamma_e^0(d) - U_{Aug}^0 - U_{SRH}^0\right)\exp\left(\frac{qV}{k_b T_c}\right)\right] \quad (10)$$

Auger and SRH equilibrium recombination rates are calculated as [58,59]:

$$U_{Aug}^0 = CNn_i^2 t \quad (11)$$

$$U_{SRH}^0 = \frac{n_i^2 t}{\tau_{SRH} N} \quad (12)$$

where $C$ is the Auger recombination coefficient taking a value of $5\times10^{-30}$ cm$^6$s$^{-1}$ for GaSb [60], $N$ is the doping concentration fixed at $10^{18}$ cm$^{-3}$ and $\tau_{SRH}$ is the lifetime associated with SRH recombination which is assumed to be 10 ns [60]. The intrinsic carrier concentration $n_i$ of GaSb is taken as $1.5\times10^{12}$ cm$^{-3}$ at 300 K [60] and the cell thickness $t$ is 10 μm. All results shown in this section are for the case of a Si emitter, since the conclusions are the same for the Drude emitter. Equivalent results for the Drude emitter are provided in Figs. S1 to S3 of Supplemental Material [61].

The impact of external luminescence on the cell *J-V* characteristic is analyzed first. The cell external luminescence has a component toward the emitter and a component toward the substrate. External luminescence towards the substrate can be minimized using efficient reflectors [62]. It follows that an external luminescence efficiency defined as:

$$\eta_{ext}(d) = \frac{\gamma^0_{e,c-e}(d)}{\gamma^0_e(d) + U^0_{Aug} + U^0_{SRH}} \tag{13}$$

where only external luminescence toward the emitter $\gamma^0_{e,c-e}$ is accounted for in the numerator reflects solely the impact of the presence of the emitter on the *J-V* characteristic. Indeed, Eq. (10) can be reformulated as:

$$J(d,V) = q\left[\gamma_a(d) - \frac{\gamma^0_{e,c-e}(d)}{\eta_{ext}(d)} \exp\left(\frac{qV}{k_b T_c}\right)\right] \tag{14}$$

and the voltage at open circuit ($J = 0$) can thus be written as [62,63]:

$$V_{oc}(d) = \frac{k_b T_c}{q} \ln\left(\frac{\gamma_a(d)}{\gamma^0_{e,c-e}(d)}\right) - \frac{k_b T_c}{q}\left|\ln[\eta_{ext}(d)]\right| = V_{oc,ideal}(d) - \frac{k_b T_c}{q}\left|\ln[\eta_{ext}(d)]\right| \tag{15}$$

where $V_{oc,ideal}$ is the open-circuit voltage in the ideal case of the radiative limit with no cell external luminescence towards the substrate. Due to the reciprocity of radiative transfer between the emitter and the cell, $\gamma_a$ and $\gamma^0_{e,c-e}$ have similar behaviors as a function of the gap thickness $d$ causing $V_{oc,ideal}$ to be relatively independent of $d$. It follows that the dependence of open-circuit voltage on gap thickness is governed by the variations of the external luminescence efficiency $\eta_{ext}$. Since $\eta_{ext} \leq 1$, Eq. (15) is written using the absolute value of its natural logarithm to emphasize that the open-circuit voltage in real conditions is effectively smaller than that of the ideal case.

External luminescence efficiency and open-circuit voltage are plotted as a function of gap thickness for different cell conditions in Fig. 6. In the radiative limit, an external luminescence efficiency of 0.5 corresponds to equal emission leaving both sides of the cell. Therefore, external luminescence towards the Si emitter when $d$ = 1000 nm is slightly affected by coherence effects of reflected propagating modes in the vacuum gap since $\eta_{ext}$ = 0.51 and evanescent modes have a negligible contribution. As the gap thickness decreases below 1000 nm, external luminescence towards the emitter becomes dominant due to tunneling of evanescent modes, such that external luminescence efficiency approaches unity and the open-circuit voltage approaches the ideal case. For a gap thickness of 10 nm, the open-circuit voltage enhancement compared to that in the far field is only 1.04 in the radiative limit, since the increase in external luminescence efficiency is only a factor of 2.09. This is because $\eta_{ext}$ = 0.45 in the far field limiting the potential enhancement factor to approximately two.

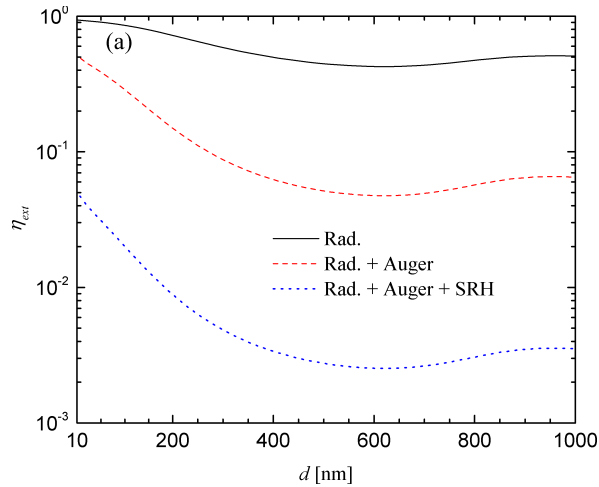

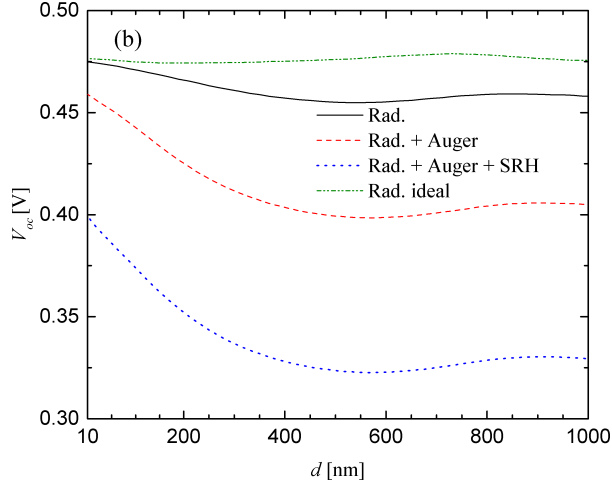

FIG. 6. (Color online) (a) Cell external luminescence efficiency ($\eta_{ext}$) and (b) open-circuit voltage ($V_{oc}$), as a function of the gap thickness (*d*) for the case of the Si emitter, when only radiative (Rad.), intrinsic (Rad. + Auger) and all (Rad. + Auger + SRH) recombination processes are considered. The open-circuit voltage in the ideal case of the radiative limit with no luminescence towards the substrate (Rad. ideal) is plotted in panel (b).

The far-field external luminescence efficiencies are $\eta_{ext}$ = 5.2×10$^{-2}$ and 2.8×10$^{-3}$, respectively, when intrinsic radiative and Auger, and when all bulk recombination processes are accounted for. This allows for significant near-field enhancement in the external luminescence efficiency and, in turn, the open-circuit voltage when non-radiative recombination mechanisms are considered. The external luminescence efficiency enhancements at *d* = 10 nm due to near-field effects are 9.84 and 18.14 leading to open-circuit voltage enhancements of 1.15 and 1.23 when intrinsic radiative and Auger, and when all bulk recombination mechanisms are considered. This impact is shown in Fig. 7, where *J-V* characteristics for the ideal case of the radiative limit with no external luminescence towards the substrate and the case when all bulk recombination is considered are plotted. The open-circuit voltage offset, $(k_b T_c / q)\left|\ln\left[\eta_{ext}(d)\right]\right|$ in Eq. (15), with respect to the ideal open-circuit voltage is much smaller at $d = 10$ nm than in the far field when all bulk recombination mechanisms are considered. Therefore, while the near-field enhancement of absorption is identical in the two cases, the near-field enhancement of open-circuit voltage is more beneficial to the cell subject to non-radiative recombination. Note that this conclusion is independent of the emitter temperature, as shown in Fig. S4 of Supplemental Material [61].

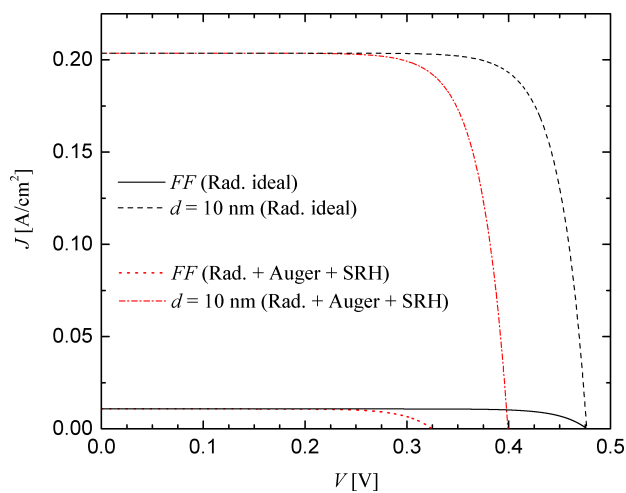

FIG. 7. (Color online) $J$-$V$ characteristics for a gap thickness of 10 nm and in the far field (*FF*) for the Si emitter, in the ideal case of the radiative limit with no luminescence towards the substrate (Rad. ideal) and that with all (Rad. + Auger + SRH) recombination processes.

The impact of gap thickness on power density at the maximum power point ($P = \max(JV)$) for the Si emitter and a 10-μm-thick cell is shown in Fig. 8. The power enhancement when $d = 10$ nm with respect to the far field case, $P/P_{FF}$, is 24.37 when considering all bulk recombination processes compared to 22.30 and 19.85 when intrinsic radiative and Auger and when only radiative recombination mechanisms are considered. The larger power enhancement when non-radiative recombination is significant is attributed to the larger enhancement in open-circuit voltage. It is concluded that TPV performance enhancement in the near field is more substantial when the cell exhibits significant non-radiative recombination. While high-quality cells always produce more power, the potential performance enhancement as the gap thickness decreases is larger for low quality cells due to the external luminescence enhancement caused by near-field effects.

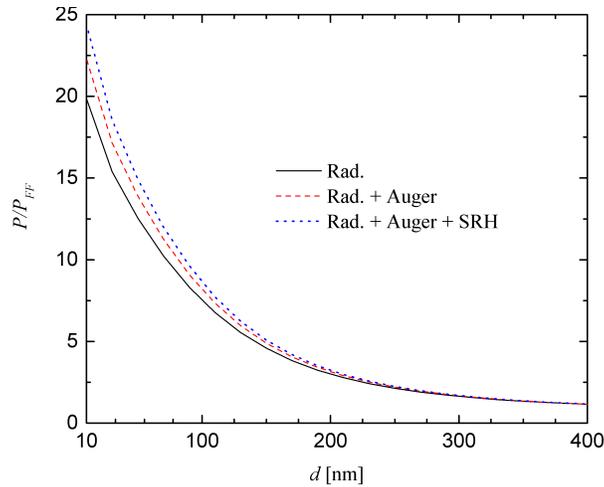

FIG. 8. (Color online) TPV power density enhancement ($P/P_{FF}$) as a function of the gap thickness ($d$) for $t = 10$ μm for the case of the Si emitter, when only radiative (Rad.), intrinsic (Rad. + Auger) and all (Rad. + Auger + SRH) recombination processes are considered.

Near-field performance enhancements at a gap thickness of 10 nm with respect to far-field conditions are summarized in Table I for the Si and Drude emitters. For the case of a Drude emitter, similar trends to that of the Si emitter are observed. The reason the near-field enhancement factors are significantly larger is because the Drude emitter is highly reflective in the far field, thus leading to an extremely small far-field external luminescence efficiency of $1.3 \times 10^{-4}$ when considering all bulk recombination mechanisms. In future work, it would be interesting to optimize the $\omega_p$ and $\Gamma$ parameters of the Drude emitter to maximize near-field TPV performance enhancement metrics, as listed in Table I, in a similar manner to Refs. [64,65] in which near-field radiative heat transfer was maximized.

TABLE I. Near-field TPV performance enhancement (power density $P$, external luminescence efficiency $\eta_{ext}$, open-circuit voltage $V_{oc}$) at a gap thickness of 10 nm with respect to far-field ($FF$) conditions. Results are given when only radiative (Rad.), intrinsic (Rad. + Auger) and all (Rad. + Auger + SRH) recombination processes are considered.

| | Silicon emitter | | |
|---|---|---|---|
| | Rad. | Rad. + Auger | Rad. + Auger + SRH |
| $P/P_{FF}$ | 19.85 | 22.30 | 24.37 |
| $\eta_{ext}/\eta_{ext,FF}$ | 2.09 | 9.84 | 18.14 |
| $V_{oc}/V_{oc,FF}$ | 1.04 | 1.15 | 1.23 |
| | Drude emitter | | |
| | Rad. | Rad. + Auger | Rad. + Auger + SRH |
| $P/P_{FF}$ | 829.39 | 1019.65 | 1238.18 |
| $\eta_{ext}/\eta_{ext,FF}$ | 26.83 | 242.45 | 562.10 |
| $V_{oc}/V_{oc,FF}$ | 1.22 | 1.45 | 1.68 |

## V. CONCLUSIONS

The near-field impacts on photon recycling and spontaneous emission rate in TPV devices were investigated. This was accomplished by analyzing the cell external luminescence, calculated via fluctuational electrodynamics, in a TPV device consisting of either an intrinsic Si or a Drude emitter and a GaSb cell. Enhanced near-field external luminescence is due to an increase in the available photon modes leading to a drop in photon recycling. The analysis showed that for gap thicknesses between the emitter and the cell less than 200 nm, evanescent modes, not contributing in far-field TPVs, dominate external luminescence for both emitter types. The enhancement of external luminescence in the near field with the Si emitter is caused by tunneling of modes propagating in the cell that otherwise fully contribute to photon recycling in far-field

TPVs. The Drude emitter supporting surface polariton modes modifies the local density of photon modes in the cell, and thus leads to an increase of the cell spontaneous emission rate. This results in an additional channel for near-field external luminescence that allows tunneling of photon modes well beyond those propagating in the cell. For gap thicknesses on the order of a few tens of nanometers, the cell near-field external luminescence with a Drude emitter can even exceed the internal luminescence of an isolated cell. For a given cell material, the impact of radiative recombination must account for the emitter type as well as the gap and cell thicknesses, which cannot be captured with a spatially uniform radiative lifetime corrected by a photon recycling factor. Finally, the impact of external luminescence on near-field TPV performance was investigated when accounting for bulk non-radiative recombination processes. The results showed that the enhancement of external luminescence in the near field, causing the cell external luminescence efficiency to increase as the gap thickness decreases, has a positive impact on the open-circuit voltage and power density. Near-field TPV performance enhancement is the largest when all non-radiative bulk recombination processes are accounted for, thus suggesting that lower quality PV cells benefit more from the near-field effects of thermal radiation than high-quality cells approaching the radiative limit.

## ACKNOWLEDGMENTS

This work was sponsored by the National Science Foundation under Grant No. CBET-1253577, and by the French National Research Agency (ANR) under Grant No. ANR-16-CE05-0013. Rodolphe Vaillon acknowledges the financial support from the College of Engineering at the University of Utah (W.W. Clyde Visiting Chair award).

## APPENDIX: TRANSMISSION FACTORS

The spectral, gap-dependent transmission factors needed to calculate the above bandgap photon flux absorbed by the cell and the above bandgap photon emission lost to the surroundings (i.e., cell external luminescence) are derived from fluctuational electrodynamics, in which a fluctuating current $\mathbf{J}^{fl}$ representing thermal emission is added to Maxwell's equations [45]. The thermally induced electric and magnetic field intensities at location $\mathbf{r}$ due to fluctuating currents contained within a heat source of volume $\upsilon$ are respectively given by:

$$\mathbf{E}(\mathbf{r},\omega) = i\omega\mu_0 \int_\upsilon \bar{\bar{\mathbf{G}}}^E(\mathbf{r},\mathbf{r}',\omega) \cdot \mathbf{J}^{fl}(\mathbf{r}',\omega) d^3\mathbf{r}' \tag{A1}$$

$$\mathbf{H}(\mathbf{r},\omega) = \int_\upsilon \bar{\bar{\mathbf{G}}}^H(\mathbf{r},\mathbf{r}',\omega) \cdot \mathbf{J}^{fl}(\mathbf{r}',\omega) d^3\mathbf{r}' \tag{A2}$$

where $\mu_0$ is the permeability of vacuum and $\bar{\bar{\mathbf{G}}}^{E(H)}$ is the electric (magnetic) dyadic Green's function relating the electric (magnetic) field with a frequency $\omega$ at location $\mathbf{r}$ to a source at $\mathbf{r}'$.

The ensemble average of the spatial correlation function of the fluctuating current is related to the local temperature of a heat source via the fluctuation-dissipation theorem [31,45]:

$$\left\langle \mathbf{J}_\alpha^{fl}(\mathbf{r}',\omega) \mathbf{J}_\beta^{fl*}(\mathbf{r}'',\omega') \right\rangle = \frac{4\omega\varepsilon_0 \operatorname{Im}(\varepsilon)}{\pi} \Theta(\omega,T,V) \delta(\mathbf{r}'-\mathbf{r}'') \delta(\omega-\omega') \delta_{\alpha\beta} \tag{A3}$$

where the subscript $^*$ refers to complex conjugate, $\varepsilon$ is the dielectric function of the heat source, $\varepsilon_0$ is the permittivity of vacuum, $\alpha$ and $\beta$ are orthogonal components ($\alpha,\beta = \rho,\theta,z$ for a polar coordinate system) indicating the state of polarization of the fluctuating current, $\delta(\mathbf{r}'-\mathbf{r}'')$ and $\delta(\omega-\omega')$ are Dirac functions, while $\delta_{\alpha\beta}$ is the Kronecker function.

The monochromatic radiative heat flux along the *z*-direction (see Fig. 1) is calculated as the ensemble average of the *z*-component of the Poynting vector [49]:

$$\langle S_z(z,\omega) \rangle = \frac{1}{2}\text{Re}\left[\langle E_\rho(z,\omega)H_\theta^*(z,\omega) - E_\theta(z,\omega)H_\rho^*(z,\omega)\rangle\right] \tag{A4}$$

The transmission factor relating the entire volume of the emitter to a discrete layer of thickness $\Delta z_j$ within the cell is obtained by calculating the difference between the flux crossing the boundary $z_j$ and the flux crossing the boundary $z_{j+1}$, and by normalizing the resulting expression by the mean energy of a generalized Planck oscillator $\Theta(\omega,T,V)$ [49]:

$$\Phi_{e-c}(\omega,d,\Delta z_j) = \frac{k_0^2}{2\pi^2}\text{Re}\left\{i\,\text{Im}(\varepsilon_e)\int_0^\infty \frac{k_\rho dk_\rho}{\text{Im}(k_{ze})}\left[\begin{pmatrix}g_{ec\rho\alpha}^E(k_\rho,z_j,\omega,d)g_{ec\theta\alpha}^{H*}(k_\rho,z_j,\omega,d)\\-g_{ec\theta\alpha}^E(k_\rho,z_j,\omega,d)g_{ec\rho\alpha}^{H*}(k_\rho,z_j,\omega,d)\end{pmatrix}\\-\begin{pmatrix}g_{ec\rho\alpha}^E(k_\rho,z_{j+1},\omega,d)g_{ec\theta\alpha}^{H*}(k_\rho,z_{j+1},\omega,d)\\-g_{ec\theta\alpha}^E(k_\rho,z_{j+1},\omega,d)g_{ec\rho\alpha}^{H*}(k_\rho,z_{j+1},\omega,d)\end{pmatrix}\right]\right\} \tag{A5}$$

where the subscript $\alpha$ involves a summation over the three orthogonal components and the *g* terms are the Weyl (plane wave) components of the dyadic Green's function [66]. These dyadic Green's functions for layered media are spatial transfer functions relating the electric and magnetic fields observed at location *z* generated by a mode with parallel wavevector $k_\rho$ and angular frequency $\omega$ to a source point located at $z'$. In Eq. (A5), an integration is performed over all possible photon modes $k_\rho$. The transmission factor relating a discrete layer $\Delta z_j$ within the cell to the entire volume of the emitter, $\Phi_{c-e}$, is equal to $\Phi_{e-c}$ due to the reciprocity of the DGFs.

The transmission factor relating a discrete layer $\Delta z_j$ within the cell to the entire volume of the substrate is given by [49]:

$$\Phi_{c-s}(\omega,d,\Delta z_j) = \frac{k_0^2}{2\pi^2}\text{Re}\left\{i\,\text{Im}(\varepsilon_s)\int_0^{k_0}\frac{k_\rho dk_\rho}{\text{Im}(k_{zs})}\left[\begin{pmatrix}g^E_{sc\rho\alpha}(k_\rho,z_{j+1},\omega,d)g^{H*}_{sc\theta\alpha}(k_\rho,z_{j+1},\omega,d)\\-g^E_{sc\theta\alpha}(k_\rho,z_{j+1},\omega,d)g^{H*}_{sc\rho\alpha}(k_\rho,z_{j+1},\omega,d)\end{pmatrix}\\-\begin{pmatrix}g^E_{sc\rho\alpha}(k_\rho,z_j,\omega,d)g^{H*}_{sc\theta\alpha}(k_\rho,z_j,\omega,d)\\-g^E_{sc\theta\alpha}(k_\rho,z_j,\omega,d)g^{H*}_{sc\rho\alpha}(k_\rho,z_j,\omega,d)\end{pmatrix}\right]\right\} \quad \text{(A6)}$$

where the summation over the photon modes $k_\rho$ is limited to propagating modes in the substrate $k_0$.

Explicit expressions for the Weyl components of the dyadic Green's functions needed to compute Eqs. (A5) and (A6) are provided in Ref. [49].

Supplemental Material for article "External luminescence and photon recycling in near-field thermophotovoltaics"

John DeSutter[*], Rodolphe Vaillon[**,*] and Mathieu Francoeur[*,†]

[*]*Radiative Energy Transfer Lab, Department of Mechanical Engineering, University of Utah, Salt Lake City, UT 84112, USA*

[**]*Univ Lyon, CNRS, INSA-Lyon, Université Claude Bernard Lyon 1, CETHIL UMR5008, F-69621, Villeurbanne, France*

(Dated: May 24, 2017)

[†] Corresponding author. Tel.: + 1 801 581 5721
Email address: mfrancoeur@mech.utah.edu

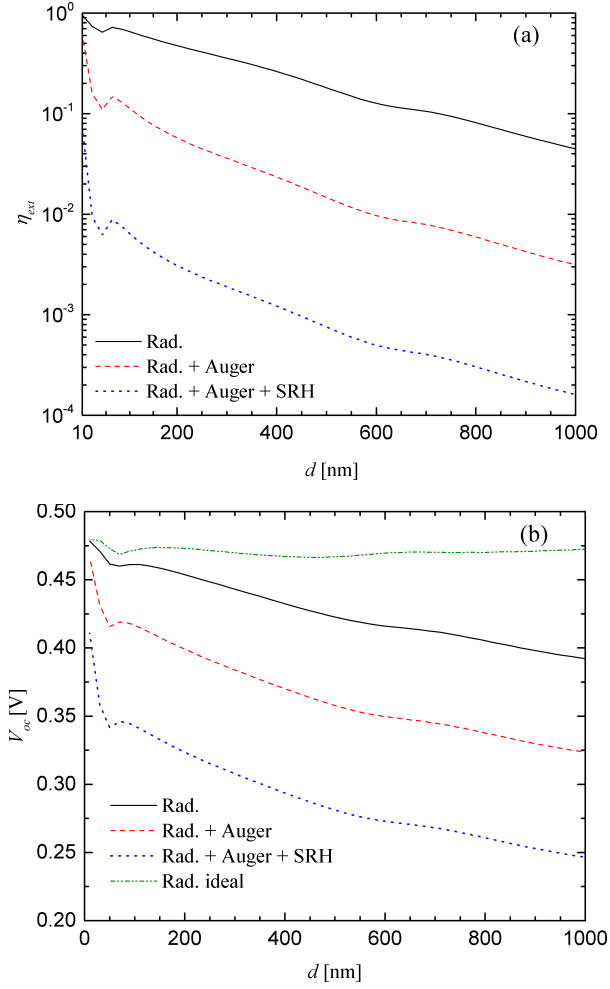

FIG. S1. (Color online) (a) Cell external luminescence efficiency ($\eta_{ext}$) and (b) open-circuit voltage ($V_{oc}$), as a function of the gap thickness ($d$) for the case of the Drude emitter, when only radiative (Rad.), intrinsic (Rad. + Auger) and all (Rad. + Auger + SRH) recombination processes are considered. The open-circuit voltage in the ideal case of the radiative limit with no luminescence towards the substrate (Rad. ideal) is plotted in panel (b).

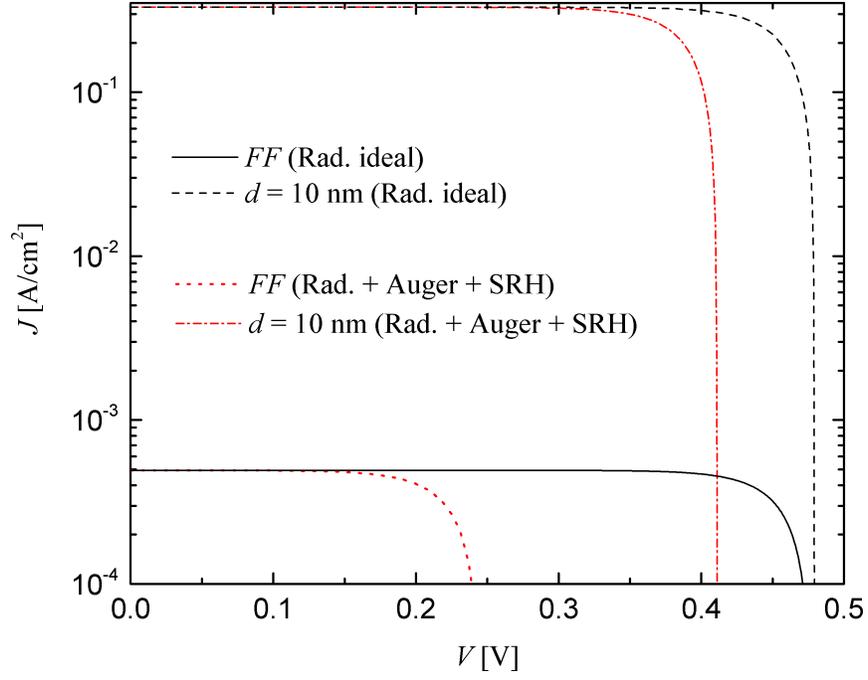

FIG. S2. (Color online) *J-V* characteristics for a gap thickness of 10 nm and in the far field (*FF*) for the Drude emitter, in the ideal case of the radiative limit with no luminescence towards the substrate (Rad. ideal) and that with all (Rad. + Auger + SRH) recombination processes.

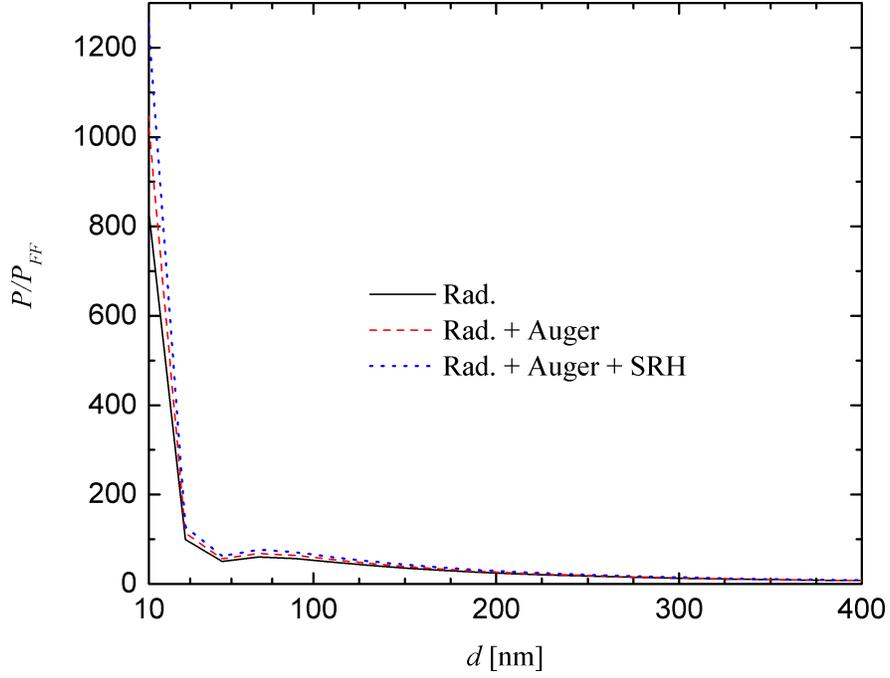

FIG. S3. (Color online) TPV power density enhancement ($P/P_{FF}$) as a function of the gap size ($d$) for $t = 10$ μm for the case of the Drude emitter, when only radiative (Rad.), intrinsic (Rad. + Auger) and all (Rad. + Auger + SRH) recombination processes are considered.

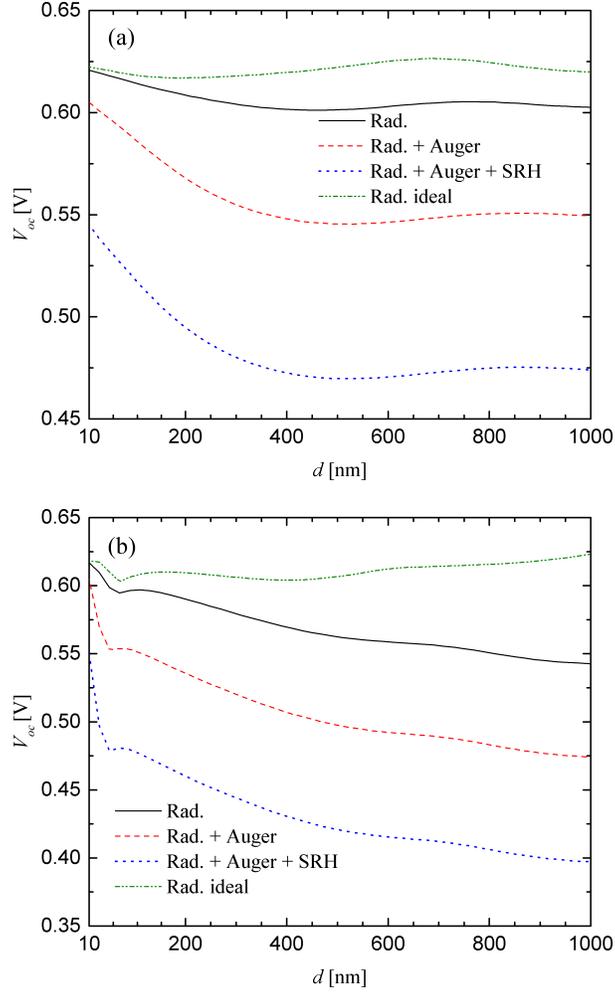

FIG. S4. (Color online) Cell open-circuit voltage ($V_{oc}$) as a function of the gap size ($d$) for $t = 10$ μm for the case of the Si (a) and Drude (b) emitters at 1500 K, when only radiative (Rad.), intrinsic (Rad. + Auger) and all (Rad. + Auger + SRH) recombination processes are considered. The open-circuit voltage in the ideal case of the radiative limit with no luminescence towards the substrate (Rad. ideal) is also plotted.

The trends for $V_{oc}$ as a function of gap size when the emitter is at 1500 K are essentially the same as those when the emitter is at 800 K (see Figs. 6(b) and S1(b)). The absolute values of $V_{oc}$ are shifted to larger magnitudes due to an increase in $V_{oc,ideal}$ as the emitter temperature increases.

This is easily identified in Eq. (15) in which the last term, $(k_b T_c / q) \left| \ln\left[\eta_{ext}(d)\right] \right|$, is unaffected by a change in emitter temperature.